\documentclass[shortnote,twocolumn]{jpsj3}
\usepackage{txfonts}
\usepackage{xspace}
\usepackage{color}

\newcommand{\EF}{$E_\mathrm{F}$\xspace}
\newcommand{\Uf}{$\mathrm{U}~5f$\xspace}
\newcommand{\Uff}{$\mathrm{U}~4f$\xspace}
\newcommand{\orb}[2]{$\mathrm{ #1 } ~ #2 $\xspace}
\newcommand{\hn}[1]{$h\nu #1~\mathrm{eV}$\xspace}
\newcommand{\EB}[1]{$E_{\mathrm{B}} #1~\mathrm{eV}$\xspace}

\newcommand{\pnt}[1]{$\mathrm{#1}$\xspace}

\newcommand{\UTe}{$\mathrm{UTe_2}$\xspace}
\newcommand{\ThTe}{$\mathrm{ThTe_2}$\xspace}
\newcommand{\UB}{$\mathrm{UB_2}$\xspace}
\newcommand{\UGe}{$\mathrm{UGe_2}$\xspace}
\newcommand{\UPd}{$\mathrm{UPd_3}$\xspace}
\newcommand{\UPt}{$\mathrm{UPt_3}$\xspace}
\newcommand{\URhGe}{$\mathrm{URhGe}$\xspace}
\newcommand{\UCoGe}{$\mathrm{UCoGe}$\xspace}

\newcommand{\etal}{\textit{et al.}\xspace}

\title{Core-Level Photoelectron Spectroscopy Study of $\mathrm{UTe_2}$}

\author{
Shin-ichi~Fujimori$^1$\thanks{fujimori@spring8.or.jp},
Ikuto~Kawasaki$^1$,
Yukiharu~Takeda$^1$,
Hiroshi~Yamagami$^{1,2}$,
Ai~Nakamura$^3$,
Yoshiya~Homma$^3$,
and Dai~Aoki$^{3}$}

\inst{$^1$Materials Sciences Research Center, Japan Atomic Energy Agency, Sayo, Hyogo 679-5148, Japan \\
$^2$Department of Physics, Faculty of Science, Kyoto Sangyo University, Kyoto 603-8555, Japan \\
$^3$Institute for Materials Research, Tohoku University, Oarai, Ibaraki 311-1313, Japan}

\abst{
The valence state of $\mathrm{UTe}_2$ was studied by core-level photoelectron spectroscopy.
The main peak position of the $\mathrm{U}~4f$ core-level spectrum of $\mathrm{UTe}_2$ coincides with that of $\mathrm{UB}_2$, which is an itinerant compound with a nearly $5f^3$ configuration.
However, the main peak of $\mathrm{UTe}_2$ is broader than that of $\mathrm{UB}_2$, and satellite structures are observed in the higher binding energy side of the main peak, which are characteristics of mixed-valence uranium compounds.
These results suggest that the $\mathrm{U}~5f$ state in $\mathrm{UTe}_2$ is in a mixed valence state with a dominant contribution from the itinerant $5f^3$ configuration. 
}

\begin{document}
\maketitle

The unconventional superconductivity in \UTe has attracted much attention in recent years \cite{UTe2_sc, UTe2_Aoki}. 
Its electronic structure is essential to understand the origin of its superconductivity, and angle resolved photoelectron spectroscopy (ARPES) has been applied using soft X-ray (SX, \hn{=565-800}) \cite{UTe2_ARPES} and vacuum ultraviolet (VUV, \hn{=30-150}) \cite{UTe2_ARPES_Miao} synchrotron radiation.
However, these two ARPES studies presented contradicting views of the electronic structure of \UTe: The SX ARPES study concluded that the band-structure calculation treating \Uf states as valence electrons can explain the overall electronic structure of \UTe \cite{UTe2_ARPES} while the VUV ARPES study argued that the near-\EF electronic structure is very similar to that of the band-structure calculation of \ThTe although there exist heavy bands around the \pnt{Z} point \cite{UTe2_ARPES_Miao}.
In addition, the partial \Uf density of states (DOS) obtained by resonant photoelectron spectroscopy (RPES) at the $4d-5f$ absorption edge (\hn{=736}) has a dominant sharp peak at the Fermi energy \cite{UTe2_ARPES}, while the on-resonant RPES spectrum measured at the $5d-5f$ absorption edge (\hn{=98}) has a dominant peak at a higher binding energy of \EB{\sim 0.7} \cite{UTe2_ARPES_Miao}.
To solve this discrepancy, additional electronic structure studies on \UTe are required.
Recently, Thomas \etal reported X-ray absorption spectrum (XAS) of \UTe under ambient and high pressures \cite{UTe2_XAS}.
They argued that \UTe exhibits intermediate valence at ambient pressure, suggesting that the \Uf state in \UTe is hybridized with the ligand states.
In the present study, we further studied the \Uf valence state of \UTe using core-level spectroscopy, which has the ability to probe the valence state of the local uranium site \cite{Ucore,SF_review_JPSJ}.
The \orb{U}{4f} spectrum of \UTe was compared with that of a typical itinerant compound, \UB, and localized compound, \UPd, as well as ferromagnetic superconductors \UGe, \UCoGe, \URhGe, and \UPt.

Photoemission experiments were conducted on the SX beamline BL23SU at SPring-8 \cite{BL23SU2}.
The overall energy resolution in the angle-integrated photoelectron spectroscopy experiments at \hn{=800} was approximately $140~\mathrm{meV}$.
The kinetic energy of photoelectrons is about $400~\mathrm{eV}$, which is considered to have enough bulk sensitivity since the \Uff spectra of $\mathrm{URu_2Si_2}$ measured at \hn{=800} \cite{Ucore,SF_review_JPSJ} and \hn{=5945} \cite{URu2Si2_HAXPES} are essentially identical.
The sample temperature was kept at $20~\mathrm{K}$ for all measurements.
Other experimental conditions are described in Ref.~\citen{UTe2_ARPES}.

\begin{figure}
	\centering
	\includegraphics[scale=0.48]{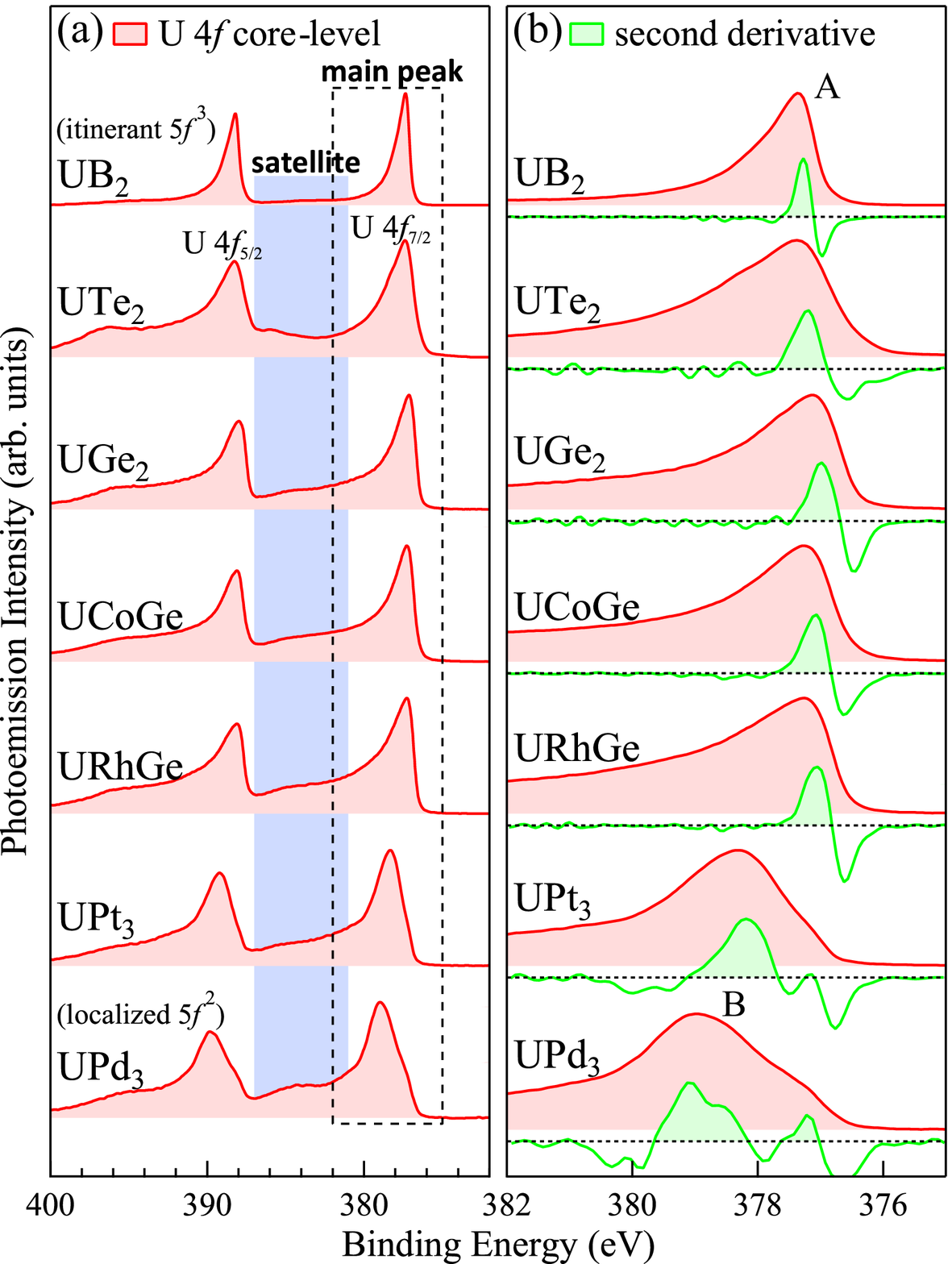}
	\caption{$\mathrm{U}~4f$ core-level spectra of \UTe and reference compounds.
Data of \UB, \UGe, \UCoGe, \URhGe, \UPt, and \UPd are depicted from Refs \citen{Ucore} and \citen{SF_review_JPSJ}.
(a) $\mathrm{U}~4f$ core-level spectra of \UB, \UTe, \UGe, \UCoGe, \URhGe, \UPt, and \UPd.
(b) Blow up of the main peaks of $\mathrm{U}~4f_{7/2}$ spectra and their negative second derivatives
}
	\label{core}
\end{figure}
Figure~\ref{core} (a) presents a comparison of the \Uff core-level spectra of \UB, \UTe, \UGe, \UCoGe, \URhGe, \UPt, and \UPd.
Their negative second derivatives of $\mathrm{U}~4f_{7/2}$ spectra are also provided in Fig.~\ref{core} (b) to indicate the locations of peaks in the spectra.
Data of \UB, \UGe, \UCoGe, \URhGe, \UPt, and \UPd are depicted from Refs \citen{Ucore} and \citen{SF_review_JPSJ}.
\UB and \UPd are typical itinerant and localized compounds, respectively.
The band structure and Fermi surface of \UB are well explained by the band-structure calculation treating all \Uf electrons as  itinerant \cite{UB2_ARPES}.
The occupation number of the \Uf state within the Muffin-Tin sphere is 2.82 in the calculation; thus, the local \Uf electronic configuration of \UB can be considered the dominant $5f^3$ configuration.
In contrast, \UPd is a uranium compound with a localized $5f^2$ configuration.

These spectra all generally consist of a dominant main peak located at \EB{=377-379} and a broad satellite structure distributed at \EB{=381-387}.
These complex spectral shape originate from the transition from the ground state to multiple final states with different local \Uf electronic configurations \cite{Fujimori_SSC}.
There are several theoretical models of the origin of the \Uff spectral profiles \cite{Okada_core,Zwicknagl_core}; however, the quantitative analysis has not yet been established.
Nevertheless, different final states have different binding energies, which can be used to identify the local electronic configuration in the ground state. 
Here, we discuss the electronic structure of \UTe based on a comparison with typical uranium compounds.

The main peak positions of \UTe, \UGe, \UCoGe, and \URhGe are almost identical (\EB{=377-377.3}), and have a similar asymmetric peak structure with a long tail toward higher binding energies.
Their main peak positions are very similar to that of the itinerant \Uf compound \UB (designated as A in Fig.~\ref{core} (b)), and are very different from the spectrum of \UPd (designated as B in Fig.~\ref{core} (b)).
This indicates that the dominant final state configurations in \UTe as well as \UGe, \UCoGe, and \URhGe are identical to that of \UB, and the dominant \Uf configurations in the ground states of \UTe, \UGe, \UCoGe, and \URhGe are also similar to that of \UB. 
In contrast, the main peaks of \UTe as well as ferromagnetic superconductors are broader than that of \UB.
As seen in the spectrum of \UPd, the main peak consists of two peaks ($E_\mathrm{B} = 378.9$ and $377.2~\mathrm{eV}$), and the broadening in the main peaks of \UTe, \UGe, \UCoGe, and \URhGe may originate from a small contribution from the \UPd-type peak on the higher binding energy side of the main peaks, although this has not been resolved experimentally.
Moreover, the core-level spectrum of \UTe is accompanied by a satellite, which has been observed in the \Uff core-level spectra of strongly-correlated or localized $5f^2$ uranium compounds.
Thus, these results indicate that the ground state of \UTe is a mixed valence state with a dominant contribution from the $5f^3$ configuration and some contribution from the $5f^2$ configuration.
These result are consistent with the result of SX-ARPES study \cite{UTe2_ARPES} and the XAS study \cite{UTe2_XAS}.
In addition, the core-level spectral shape of \UTe is similar to that of \UGe, \UCoGe, and \URhGe, which have essentially itinerant but correlated \Uf states \cite{URhGe_ARPES,UGe2_UCoGe_ARPES,SF_review_JPSJ}, suggesting that \UTe should be similar to them.

Here, we consider the relationship between the present result and the results of other studies on the electronic structure of \UTe.
In density functional theory (DFT) plus Hubbard $U$ (DFT+$U$) and generalized gradient approximation plus $U$ (GGA+$U$) with $U \gtrsim 2~\mathrm{eV}$, quasi-two-dimensional Fermi surfaces have been predicted \cite{UTe2_Xu,UTe2_Ishizuka}.
In these calculations, most of the \Uf weight was away from the Fermi level by the introduction of the $U$, and the topology of the Fermi surface becomes almost identical to that of the DFT calculation for \ThTe.
Experimentally, the VUV ARPES study reported very similar near-\EF electronic structure, although the existence of a heavy band around the $\mathrm{Z}$ point was claimed \cite{UTe2_ARPES_Miao}.
Furthermore, the VUV-RPES spectrum was interpreted based on the ground state with the dominant $5f^2$ Hund's rule ground state, which is based on the slightly mixed valent but essentially localized $5f^2$ state \cite{UTe2_ARPES_Miao}.
In such situation, its core-level spectrum should be similar to those of the localized compound \UPd or weakly hybridized compound \UPt.
However, the present result indicates that the hybridized (itinerant) $5f^3$ configuration is dominant in the ground state of \UTe, and the \Uf states should thus make dominant contributions to the state at the Fermi level.
The very different nature of \Uf states observed in the VUV ARPES study may originate from the enhanced surface sensitivity of VUV PES ($\lesssim 5~\mathrm{\AA}$) compared with SX-PES ($\gtrsim 15~\mathrm{\AA}$), as similar discrepancies have been observed in strongly correlated $f$-electron materials \cite{US_PES,SF_review_JPSJ,SF_review_JPCM}.

In summary, we applied core-level spectroscopy to \UTe.
A comparison between the core-level spectral shape of \UTe and that of typical compounds demonstrated that the local electronic configuration of the \Uf state in \UTe is in the mixed valence state with a dominant contribution from the $5f^3$ configuration.
Furthermore, the spectrum of \UTe is very similar to that of \UGe, \UCoGe, and \URhGe, suggesting that \Uf should essentially have  itinerant character, although there exist electron correlation effects.
The result indicates that the topology of the Fermi surface of \UTe should be considerably different from the localized model, such as the DFT calculation for \ThTe.

\begin{acknowledgment}
The authors thank A. B. Shick and W. E. Pickett for stimulating discussion. 
The experiment was performed under Proposal Nos. 2019A3811 at SPring-8 BL23SU.
The present work was financially supported by JSPS KAKENHI Grant Numbers JP15H05882, JP15H05884, JP15H05745, JP15K21732, JP16H01084, JP16H04006, JP18K03553, and JP19H00646.
\end{acknowledgment}

\bibliographystyle{jpsj}
\bibliography{25523}

\end{document}